# A Novel String Distance Function based on Most Frequent K Characters

Sadi Evren Seker, Oguz Altun, Uğur Ayan and Cihan Mert

*Abstract*—This study aims to publish a novel similarity metric to increase the speed of comparison operations. Also the new metric is suitable for distance-based operations among strings.

Most of the simple calculation methods, such as string length are fast to calculate but doesn't represent the string correctly. On the other hand the methods like keeping the histogram over all characters in the string are slower but good to represent the string characteristics in some areas, like natural language.

We propose a new metric, easy to calculate and satisfactory for string comparison.

Method is built on a hash function, which gets a string at any size and outputs the most frequent K characters with their frequencies.

The outputs are open for comparison and our studies showed that the success rate is quite satisfactory for the text mining operations.

## I. Introduction

Most of the string distance functions are based on the character sequence of the string. Besides the similarity of characters, the order of characters is considered to be important in most of the string similarity metrics. By the impact of big data studies, the time and memory complexity of the string similarity metrics are considered to be more important.

We propose a new string similarity metric, which is built over a hashing function.

In this paper, we will briefly introduce the idea behind string similarity metrics and their applications. After the idea of string similarity, we will introduce some of the advanced hashing functions and their suitability on string metrics.

Finally we will introduce a novel string similarity metric and we will discuss the success rate of novel method over the current methods.

## II. String Distance Functions

The string distance functions or string similarity metrics are defined between two strings, let's say str1 and str2. The function can be defined as a relation from a domain to range.

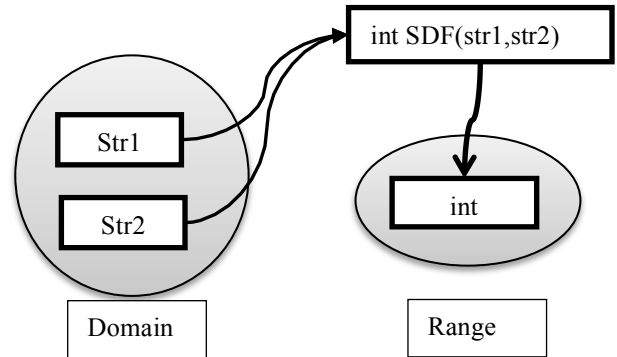

Fig. 1. Generic view of a SDF

Most of the time, the function is with two parameters where both of them are strings and the return value is an integer.

The generic view of a String Distance Function (SDF) is demonstrated in Fig. 1.

All the SDF implementations can be considered as a hash function where the function is working in one direction and the output is keeping less memory space.

For example levenshtein-distance [1] is a function which gets two parameters and calculates the edit distance between two strings. The three operations, delete, insert or update over the characters of a string are considered as an edit and each edit can be scored depending on the implementation.

$$lev_{a,b}(i,j) = \begin{cases} \max(i,j) & \text{if } \min(i,j) = 0, \\ \min \begin{cases} lev_{a,b}(i-1,j)+1 \\ lev_{a,b}(i,j-1)+1 \\ lev_{a,b}(i-1,j-1)+[a_i \neq b_j] \end{cases} & \text{otherwise.} \end{cases} \quad (1)$$

Finally a score of integer is collected from the SDF and the function is irreversible from the output integer to the initial strings.

On the other hand the output is in integer form which keeps less memory space than the input strings.

Some other methods like Tanimoto Distance [2] or Jaccard Coefficient [3] is built on the bitwise operators. In these methods the strings are considered in the bit level (not the character level as in Levenshtein Distance) and the number of equality in the bit level are considered as a score of similarity.

$$T_s(X,Y) = \frac{\sum_i (X_i \wedge Y_i)}{\sum_i (X_i \vee Y_i)} \quad (2)$$

Tanimoto Distance for example, sums the bit level 'and' and 'or' operations and divides these summations to get the similarity.

The distance function is the logarithm of this similarity.

$$T_d(X,Y) = -\log_2(T_s(X,Y)) \quad (3)$$

Also Jaccard coefficient or Jaccard Index is based on the similar methods where the 'and' and 'or' operations are replaced with 'set intersection' and 'set union' operations.

$$J(A,B) = \frac{|A \cap B|}{|A \cup B|} \quad (4)$$

And the distance function can be calculated by subtracting this value from 1.

$$d_J(A,B) = 1 - J(A,B) = \frac{|A \cup B| - |A \cap B|}{|A \cup B|} \quad (5)$$

Another SDF called Hamming Distance [4] is based on the matching and mismatching characters in the order of strings.

The bit level distance can be represented as a hypercube for the Hamming Distance as in Fig. 2.

In the Hamming Distance, any letters, which do not match each other, are considered as 1 and the summation of those mismatches are considered as the distance between two strings.

We can summarize the SDFs in two groups. First group is really fast and good in memory but the outputs are meaningless for the natural language string comparisons.

The second SDF group is quite satisfactory on the natural language string comparisons but their time complexity is high.

For example, Hamming Distance is in first SDF group with really good, low time complexity but the strings are considered far away even their meanings are close to each other.

Consider the example of 'revolution' and 'evolution' where the first word is derived from the second word and the distance between two words is 9, which means they are completely unrelated words. The same problem occurs for bitwise comparisons like Tanimoto or Jaccard SDFs.

On the other hand a good SDF like levenshtein distance can find the similarity between words 'revolution' and 'evolution' as 1 since there is only 1 letter deleted from first to second, but this operation will take much more time than the previous functions.

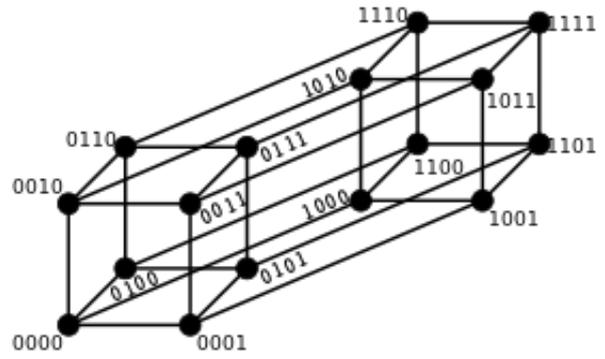

Fig. 2. Generic view of Hamming Distance hypercube

Although there are dynamic programming approaches to reduce the time complexity of the functions like Levenshtein Distance, those implementations increases the memory complexity of the algorithm.

In this study we propose an alternative SDF for comparing two strings with a better time complexity than Levenshtein Distance and a higher satisfactory distance metric than Hamming Distance.

### III. STRING HASHING ALGORITHMS

In the essence the SDFs can be considered as a hash function defined on two separate strings. The SDF function can be considered as a trapdoor function, where there is no turn back from output to input (irreversible). Also the SDF output is a summary of the differences between two strings, where it is most of the time symbolized as an integer.

The one of most widely used hashing function group is substitution permutation network (SPN) [5]. In this hashing method, the input is considered as a plain text and the plain text is processed through the steps with 'permutation', 'substitution', 'exclusive or' or 'splitting' until reaching the hashed text.

The generic view of the SPN hashing method is demonstrated in Fig. 3.

Also another mostly implemented method is using building networks on bitwise operations like message digest 5 (MD5) algorithm [6] does.

In Fig. 4, the generic view of MD5 hashing is demonstrated. In each iteration the input text is divided into 4 parts and only one of the four parts (which is A in the demonstration) is subjected to the bitwise operations with the rest 3 parts of the input.

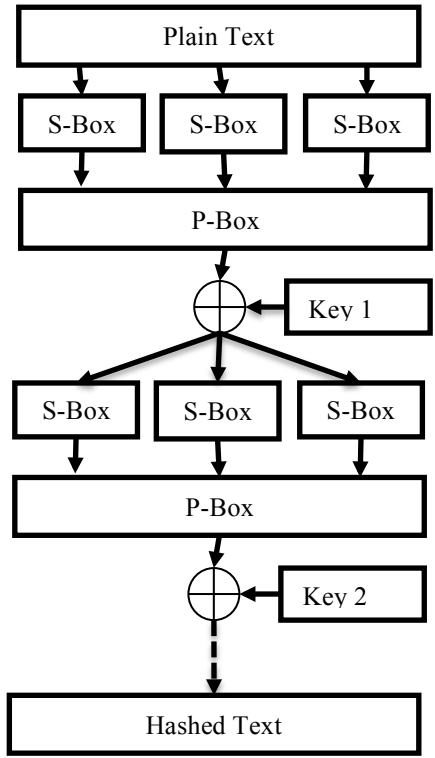

Fig. 3. Generic view of a SPN Hashing

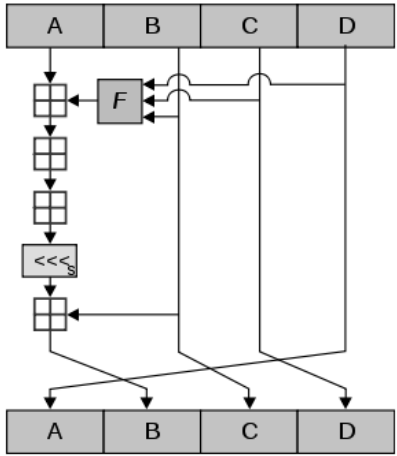

Fig. 4. Generic view of a MD5 Hashing

A similar approach is applied for most of the hashing algorithms. For example SHA1 algorithm follows a similar bitwise level operation in its implementation.

Besides the above bitwise hashing algorithms, there is another group of hashing which is mostly accepted as primitive hashing functions.

In this group of hashing, the strings are manipulated with the primitive operations like truncating or getting character frequency. These implementations can be considered as a simpler way and the results can be predicted by human much more easily.

For example, even getting the first letter of each string can be considered as a hashing function. Some hashing functions gets the certain letters like first, third and last to find out the hashed version. Also transforming an input string while keeping some part of its semantic is another important issue in the natural language processing (NLP) studies [7].

For example part of speech taggers (POS-Tagger) or stemmers can be considered in this group where they keep some semantic information on the output [8].

### IV. A NOVEL STRING SIMILARITY METRIC

This section describes the similarity metric proposed in this study.

Any string is processed through the hash function which outputs the most frequent two characters. Sorting and getting the most frequent characters and the number of occurrences can achieve this operation.

Our SDF study can be divided into two parts. In the first part, the hashing function maximum frequent two is applied over the both of the input strings.

**Algorithm 1: MaxFreq2Hashing**

1. $X \leftarrow h(str)$
2. for $i \leftarrow 0$ to length(str1)
3.    putHashMap(str_i, count(getHashMap(str_i)+1)
4.    $c1 \leftarrow getChar(maxHashMap,1)$
5.    $n1 \leftarrow getCount(maxHashMap,1)$
6.    $c2 \leftarrow getChar(maxHashMap,2)$
7.    $n2 \leftarrow getCount(maxHashMap,2)$
8.    $x1 \leftarrow concat(c1,n1,c2,n2)$
9.    return x1

In the maximum frequency hashing algorithm, we output a string of length 4 where the first and third elements keep the characters and second and fourth elements keep the frequency of these characters. If the frequency of two characters in the string is equal, the first occurrence of the character is returned.

In the case of all frequencies of a character in string is equal to each other, than the hashing function works like returning the first two characters.

On the second part of SDF, the hashed output of the strings is compared with the algorithm 2.

**Algorithm 2: Novel SDF**

1. Let str1 and str2 be two strings to measeure the distance between
2. $X \leftarrow f(str1,str2,limit)$
3.    $x1 := h(str1)$
4.    $x2 := h(str2)$
5.    def similarity :=0
6.    if $x1[0]==x2[0]$ then
7.      similarity := similarity + x1[1]+x2[1]
8.    if $x1[0]==x2[2]$ then
9.      similarity := similarity + x1[1]+x2[3]
10.   if $x1[2]==x2[0]$ then
11.     similarity := similarity + x1[3]+x2[1]
12.   if $x1[2]==x2[2]$ then
13.     similarity := similarity + x1[3]+x2[3]
14.   retun limit-similarity

Execution of SDF function will return a real number between 0 and limit. By default in our studies we have taken limit as 10 since we don't want a minus distance value and the possibility of 10 occurrence of the two maximum frequency characters common between two strings is low. If the output of the function is 10 we can interpret the case as there is no common character and any value below 10 means there are some common characters shared by the strings.

**Sample Run**

Let's consider maximum 2 frequent hashing over two strings 'research' and 'seeking'.

h('research') = r2e2

because we have 2 'r' and 2 'e' characters with the highest frequency and we return in the order they appear in the string.

h('seeking') = e2s1

Again we have character 'e' with highest frequency and rest of the characters have same frequency of 1, so we return the first character of equal frequencies, which is 's'.

Finally we make the comparison:

TABLE I: SAMPLE RUNS WITH HASHING STEPS

|  | Hashing Outputs | SDF Output |
|---|---|---|
| 'night' 'nacht' | n1i1 n1a1 | 9 |
| 'my' 'a' | m1y1 a1NULL0 | 10 |
| 'research' 'research' | r2e2 r2e2 | 6 |
| 'aaaaabbbb' 'abababab' | a5b4 a5b4 | 1 |
| 'significant' 'capabilities' | i3n2 i3a2 | 5 |

f('seeking','research',10) = 8.

We simply compared the outputs and only the number of 2 and result is 10-8 = 2.

Table I holds some sample runs between example inputs.

In all above cases, the limit value is assigned as 10. The function can also be implemented for the any string like binary numbers or nucleotide sequences.

In binary numbers case, the function works exactly same as comparing the number of 1s and 0s in both string.

In genetic area, the function can work with the limit value of maximum string length. For example two partial strings in FASTA format can be compared as below:

Str1=

LCLYTHIGRNIYYGSYLYSETWNTGIMLLLITMATAF MGYVLPWGQMSFWGATVITNLFSAIPYIGTNLV

Str2 =

EWIWGGFSVDKATLNRFFAFHFILPFTMVALAGVHLT FLHETGSNNPLGLTSDSDKIPFHPYYTIKDFLG

h(str1) = L9T8

h(str2) = F9L8

f(str1,str2,100) = 83

Experiments holding.

## V. EXPERIMENTS

This section explains the methodology of experiments run over the IMDB62 data set and the classification methods applied after the feature extraction methods. In this study two different feature hashing method is directly applied over the plain text.

  i)   Levenshtein Distance
  ii)  Jaccard Index
  iii) MaxKFreqHashing

This study compares the success rate and running time of the above methods.

Finally the evaluations of feature hashing methods are applied on the author recognition via the classification algorithms, k-nearest neighborhood (KNN) [9]. The results are evaluated via the root mean square error (RMSE) [9] and relative absolute error (RAE) [10].

### A. Dataset

We have implemented our approach onto IMDB62. Table II demonstrates the features of the datasets. In the IMDB62 database, there are 62 authors with a thousand of comments for each of the authors. The database is gathered from the internet movie database[1] which is available for the authors upon request.

TABLE II: SUMMARY OF DATASET

|  | IMDB62 |
|---|---|
| Authors | 62000 |
| Texts per Author | 1000 |
| Average number of words per entry | 300 |
| Std. Dev. of words per author | 198 |
| Number of distinct words in corpus | 139.434 |

The dataset is quite well formed for the research purposes. Unfortunately in a plain approach to text mining, like word count, the hardwares in the study environment would not qualify the requirements for the feature extraction of all the terms in data source which is 139,434 for IMDB data set.

*Memory Requirement = 139,434 words x 62,000 posts x 300 average word length x 2 bytes for each character = ~ 4830 GByte*

The amount required to process the data set via the word counts requires a feature vector, allocating memory for each of the distinct words. After applying the feature hashing methods, the number of bits required can be reduced to quite processable amount. For example, in the novel hashing method, we propose, the number of bits is reduced to 16.

### B. Execution

In the execution phase, we have implemented a word tokenizing over the data set. Each author has a feature vector of words.

We have applied the ensemble classification [11] over the classification algorithms KNN, SVM and ANN where they run over the feature vectors to classify the texts between authors.

The success rate is calculated by the percent of correctly classified texts between authors during the test runs.

The training and test data sets division is done by the 10-fold cross validation method, where a never used 10% of data set is spared for the testing and rest 90% is taken in the trainin phase for 10 runs.

---

[1] IMDB, internet movie database is a web page holding the comments and reviews of the users and freely accessible from www.imdb.com address.

*B. Results*

During the execution some parameters effect the success rate and running performance. We have specially concentrated on the K parameter in MaxKFreqHashing algorithm, which is in the core of novel distance metric.

Increasing the K value effects the success rate. The increase on the success rate is demonstrated in Fig. 5 and the increase is meaningless after parameter 3, since the success goes uf from 65% to 68% which is omittable in this study.

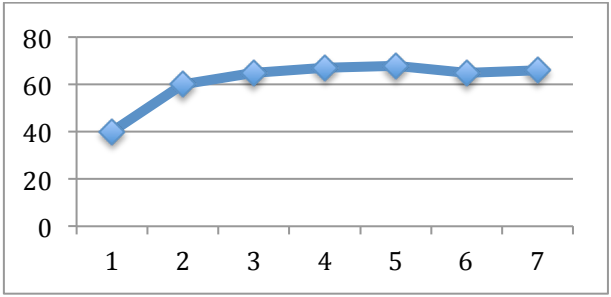

Fig. 5. Effect of K parameter on the success rate for MaxKFreqHashing

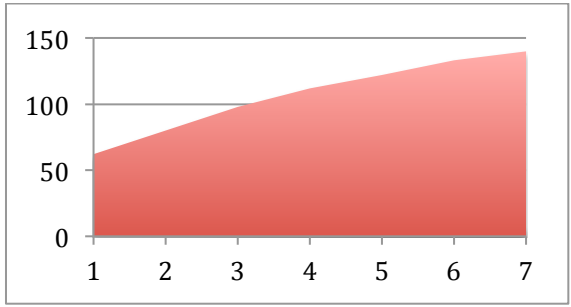

Fig. 6. Effect of K parameter on running time performance.

The effect of K parameter is demonstrated on Fig. 6. Please note that, after K value 5, there is a slight decrease on the success rate. Reason of this decrease is the increasing number of nulls for the short length words. Increasing the K value also effects the time performance of the algorithm.

The increase of K parameter increases the running time of the algorithm. Also the size of feature vector, after the execution of hashing algorithm increases if the K value increases.

The distance metrics works after the execution of hashing algorithm.

The low error rate on Table III indicates a higher success. During the comparison of methods, in the Table III, k parameter of novel SDF is 2.

Novel SDF proposed in this paper has a slightly worse success rate than levenshtein distance but it is better in the running time.

The performance of each distance function with the parameters in Table III is given in Table IV.

The time complexities are calculated with keeping the memory complexity O(n+m), where n is the string length of first string and m is the length of second string. There are better time complexity functions for Levenshtein Distance and Novel SDF with dynamic memory implementation.

The calculation of time complexity of novel SDF is quite simple. In order to get the maximum frequent K characters from a string, the first step is sorting the string in a lexiconical manner. After this sort, the input with highest occurance can be achived with a simple pass in linear time complexity. Since major classical sorting algorithms are working in O(nlogn) complexity like merge sort or quick sort, we can sort the first string in O(nlogn) and second string on O(mlogm) times. The total complexity would be O(nlog n ) + O (m log m) which is O(n log n) as the upper bound worst case analysis.

VI. CONCLUSION

In this paper a novel string distance function has been proposed. The function is built on two steps, in the first step the maximum frequent K characters are gathered with their frequencies from the string. In the second step, the hash results from first step is calculated in a special way and the distance between two strings are calculated.

TABLE III: ERROR RATES OF DISTANCE METHODS

|  | RMSE | RAE |
|---|---|---|
| **Levenshtein Distance** | **29** | **0.47** |
| Jaccard Index | 45 | 0.68 |
| *Novel SDF* | *32* | *0.49* |

TABLE IV: CUMULATIVE RUNNING TIMES

|  | Running Time | Time Complexity |
|---|---|---|
| Levenshtein Distance | 3647286.54sec | O(n*m) = O(n²) |
| Jaccard Index | 228647.22sec | O(n+m)= O(n) |
| Novel SDF | 2712323.51sec | O(nlog n+mlog m) =O(nlog n) |

The novel string distance function has been tested on a real world natural language data set for author recognition problem and yielded a better result than Jaccard index and run faster than levenshtein distance with k=2 parameter setting.

By the success rate and time performance, we can claim the novel string distance function is quite faster than full text analysis functions like levenshtein distance, pos tagging or tf-idf and much more successful than the bitwise operating string distance functions like Jaccard index, Tanimoto Distance or Hamming distance.

We believe this novel string distance function will be useful in many areas like bioinformatics, natural language processing or text mining.

REFERENCES

[1] Levenshtein VI (1966). "Binary codes capable of correcting deletions, insertions, and reversals," *Soviet Physics Doklady*, 10: 707–10.
[2] J. D. Rogers and T. T. Tanimoto, "A Computer program for classifying plants". *Science*, 132(3434): 1115–1118. doi:10.1126/science.132.3434.1115, 1960.
[3] P. Jaccard, "The distribution of the flora in the alpine zone*", New Phytologist* 11: 37–50, 1912.
[4] W. R. Hamming, "Error detecting and error correcting codes", *Bell System Technical Journal*, 29 (2): 147–160, MR 0035935.1950.
[5] S. E. SEKER and C. Mert, "A novel feature hashing for text mining," *Journal Of Technical Science And Technologies*, 2(1), 2013, pp 37-40.
[6] R. Rivest, "The MD5 message-digest algorithm," *Internet RFC 1321*, April 1992.
[7] S. E. SEKER, B. DIRI, "TimeML and Turkish temporal logic", International Conferenc on Artificial Intelligence, 2010, IC-AI'10, pp. 881-887

[8] S. E. SEKER, Z. Erdem, N. Ozalp, C. Mert, K. Al-Naami, "Correlation between Turkish stock market and economy news", *International Workshop on Relaibility Aware Data Fusion in Participatory Networks*, May 2013, Austin Texas.
[9] I. Ocak and S. E. SEKER, "Estimation of elastic modulus of intact rocks by artificial neural network," *Rock Mechanics and Rock Engineering, Springer*, DOI: 10.1007/s00603-012-0236-z, 2012.
[10] I. Ocak and S. E. SEKER, "Calculation of surface settlements caused by EPBM tunneling using artificial neural network, SVM, and Gaussian processes," *Environmental Earth Sciences, Springer-Verlag*, DOI: 10.1007/s12665-012-2214-x, 2013.
[11] S. E. SEKER, C. Mert, K. Al-Naami, U. Ayan, N. Ozalp, "Ensemble classification over stock market time series and economy news", *Intelligence and Security Informatics (ISI), 2013 IEEE International Conference on*, 2013, pp. 272-273

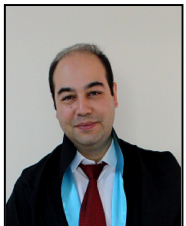

**Sadi Evren SEKER** was born in Istanbul in 1979. He has completed his BSc., MSc. and PhD. Degrees in computer science major. He also holds an M.A. degree in Science Technology and Society. His main research areas are Business Intelligence and Data Mining. During his post-doc study, he has joined data mining research projects in UTDallas. He is currently Asst. Prof. in Istanbul Medeniyet University, Department of Business and studying on data mining and business intelligence topics. He is an IEEE member and senior member of IEDRC. He has more than 20 peer-reviewed papers published or accepted in last year.

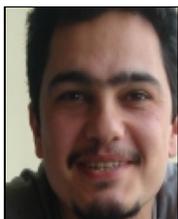

**Oguz Altun** was born in Kahramanmaras, Turkey in 1974. He got his Bachelor in Physics from Middle East Technical University, Ankara, Turkey in 1997, Master in Computer Science from University of Chicago, Chicago, USA in 2001, and PhD in Computer Engineering from Yildiz Technical University in Istanbul, Turkey in 2010.

He worked as a research assistant in Yildiz Technical University between 2001 and 2010, and started postdoc work in 2011 in Computer Engineering Department of Epoka University, Albania, where he still works as a faculty member and department head. His research interests include most artificial intelligence subjects, especially computer vision (in which he did his PhD dissertation) and metaheuristic optimization. He is co-author of numerous papers in scientific journals and conferences in these fields.

Dr. Altun is the editor of the INISTA 2007 conference proceedings book and chair of the ISCIM 2013 conference.

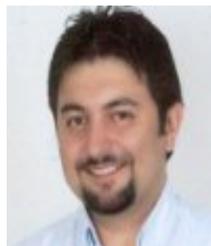

**Ugur Ayan** has worked as a Director in Informatics and Information Security Research Center (BILGEM) of The Scientific and Technological Research Council of Turkey (TUBITAK). He obtain his B.S. Diploma in Engineering Faculty (2001) from Bogazici University, Turkey; M.S. degree in System and Control Engineering (2004) from Bogazici University and his Ph.D. in Faculty of Electrical & Electronics Engineering (2010) from Yildiz Technical University, Istanbul, Turkey. He worked as a researcher at Computational Physiology Lab, University of Houston (under supervised of Eckhard Pfeiffer Professor Ioannis Pavlidis).

Dr. Ayan has worked as a Director at TUBITAK BILGEM since October, 2012. He worked as a senior researcher at TUBITAK BILGEM from September, 2010 to October, 2012. He is also an adjunct Assistant Professor at Computer Science & Engineering Department of Vistula University, Warsaw, Poland. Before joining BILGEM, he has worked as an lecturer, instructor, teaching and research assistant at Computer Engineering Department of the Turkish Air Force Academy, Istanbul Kultur University and Halic University, Istanbul, Turkey, about 10 years. His research interests are based on data mining in huge data, signal processing, bioinformatics and machine learning algorithms. He has written several articles on bioinformatics and data mining.

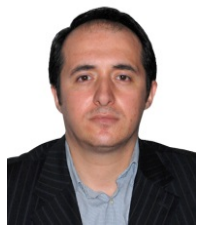

**Cihan Mert** joined The Computer Technologies and Engineering Faculty (CTEF) of International Black Sea University (IBSU) in 2011 after nearly 15 years of service to mathematical education. His educational background is as follows: B.S., Middle East Technical University, Education Faculty, Teacher of Maths, Ankara/Turkey, 1997, M.Sc., Georgian Technical University, Faculty of Informatics and Control Systems, Tbilisi/Georgia, 2000, and Ph.D., International Black Sea University, the Faculty of Computer Technologies and Engineering, Tbilisi/Georgia, 2011.

Before joining to IBSU, he worked as a Teacher of Math from 1997 to 2011 at the same time he served as the Head of Math Department. Then in 2011 He's joined IBSU and became lecturer and the Head of Master Studies in CTEF, IBSU. Then he was visiting scholar at University of Texas at Dallas from August, 2012 to August, 2013. Now he is the Dean of CTEF, IBSU. His research interests are time series analysis, signal processing, and Data mining. He has several articles on time series analysis and data mining.